\documentclass[twocolumn, authoryear]{aastex631}
\usepackage{CJK}
\usepackage{amsmath,amsthm,amssymb,amsfonts,booktabs}

\newcommand{\kepler}[0]{\emph{Kepler}}

\newcommand{\tess}[0]{\emph{TESS}}

\newcommand{\teff}[0]{$T_{\text{eff}}$}

\newcommand{\Dnu}[0]{$\Delta\nu$}
\newcommand{\numax}[0]{$\nu_{\mathrm{max}}$}
\newcommand{\muhz}[0]{$\mu{\mathrm{Hz}}$}

\begin{document}
\begin{CJK*}{UTF8}{gbsn}

\title{Asteroseismology with TESS: Emergence of Dipole Mode Suppression From Subgiants?}

\author[0009-0000-5381-7039]{Shurui Lin~(林书睿）}
\affiliation{Department of Astronomy, School of Physical Sciences, University of Science and Technology of China, Hefei, Anhui 230026, China}
\affiliation{CAS Key Laboratory for Researches in Galaxies and Cosmology, School of Astronomy and Space Science, University of Science and Technology of China, Hefei, Anhui 230026, China}
\affiliation{Department of Astronomy, University of Illinois at Urbana-Champaign, 1002 West Green Street, Urbana, IL 61801, USA}
\author[0000-0001-6396-2563]{Tanda Li~（李坦达）}
\affiliation{Institute for Frontiers in Astronomy and Astrophysics, Beijing Normal University, Beijing 102206, China}
\affiliation{School of Physics and Astronomy, Beijing Normal University, Beijing, 100875, Peopleʼs Republic of China}
\affiliation{School of Physics and Astronomy, The University of Birmingham, UK, B15 2TT}
\author[0000-0001-8317-2788]{Shude Mao~（毛淑德）}
\affiliation{Department of Astronomy, Tsinghua University, Beijing 100084, China}
\author[0000-0002-4544-0750]{Jim Fuller}
\affiliation{TAPIR, Walter Burke Institute for Theoretical Physics, Caltech, Mailcode 350-17, Pasadena, 91125, CA, USA}

\correspondingauthor{Shurui Lin (shuruil3@illinois.edu), Tanda Li (litanda@bnu.edu.cn)}

\begin{abstract}
Dipole mode suppression is an observed behavior of solar-like oscillations in evolved stars. This study aims to search for depressed dipole modes in giant stars using data from the Transiting Exoplanet Survey Satellite (TESS) and investigate when the suppression starts to emerge. 
We study a sample of 8,651 TESS-evolved stars and find 179 stars with significant dipole mode depression by comparing the oscillation amplitudes of radial and dipole modes. Notably, 11 of them are located near the base of the red-giant branch, indicating that mode suppression appears earlier than the point inferred in previous studies with the Kepler data. These findings provide new evidence for the dipole mode suppression in giant stars, particularly in subgiants.
\end{abstract}


\section{Introduction}
\label{sec:intro}

The phenomenon of depressed oscillations in red giant stars has attracted considerable attention due to its potential to reveal the internal dynamics of these stars. \cite{Fuller2015Science} proposed the``magnetic greenhouse effect'', suggesting that a magnetic field could trap gravity waves within the radiative core, leading to the dissipation of oscillations and the suppression of mixed modes. Later, analyses of Kepler data revealed that approximately 20\% of red giants exhibit depressed dipole mode oscillations, particularly in stars with masses above 1.1 solar masses. Remarkably, this phenomenon was observed in 60\% of stars with masses between 1.6 $M_\odot$ and 2.0 $M_\odot$ \citep{Stello2016Nature}. 
Later, quadrupole mode suppression was also found in the Kepler data, as predicted by the magnetic theory \citep{Stello2016pasa}.
While the internal magnetic field is a plausible explanation for mode suppression, it may not fully account for the partial suppression observed in some stars, suggesting that additional mechanisms could be at play.
\citet{Mosser2017AA} found that the depressed dipole modes in red giants are mixed modes, rather than pure pressure modes and that their global seismic properties remain similar to those of normal stars. However, the magnetic field does not fully align with this constraint, so the mechanism behind dipole mode suppression still warrants further investigation.

The magnetic origin of depressed dipole modes offers a new avenue for studying stellar evolution, as magnetic fields significantly influence the angular momentum distribution within stars, thereby affecting stellar rotation and the internal transport of elements \citep{Fuller2019MNRAS, JFD2009ARAA}. Various dynamical processes could generate the stellar magnetic field. For instance, in stars with convective cores, the $\alpha$-$\omega$ effect can create and amplify magnetic fields \citep{alpha-omega}. Inherent dynamical instabilities in differentially rotating stars can also induce internal magnetic fields \citep{Mag_dynamo}. Additionally, these magnetic fields could be remnants from the molecular clouds that formed the stars, as supported by numerical simulations suggesting that such “fossil fields” might persist throughout stellar evolution \citep{Ferrario2015}. Furthermore, simulations indicate that stellar mergers could produce massive stars with significant internal magnetic fields \citep{Schneider2019Nature}. Each of these potential mechanisms awaits further observational validation.

In addition to the aforementioned study of red giants, \citet{Garcia2014AAp} found depressed dipole mode in late subgiant KIC 8561221 (with $T_\mathrm{eff} = 5281$ $\mathrm{K}$ and \numax{} = 490 $\mu\mathrm{Hz}$) with the help of short cadence data ($58.85$ s) of Kepler, echoing a hint for mode suppression in early giant stars. However, early red giants and subgiants can not be fully examined for depressed dipole mode oscillations due to the limit of the Nyquist frequency of Kepler long-cadence data ($273.16$ $\mu \mathrm{Hz}$) and the limited availability of short-cadence observations. This highlights the need for more short-cadence data of early-type stars to understand the origin of mode suppression better.

In this paper, we explore whether early-stage giant branch stars exhibit suppressed dipole mode oscillations using TESS data, which includes 2-minute and 20-second cadence observations. By analyzing these younger stars, we aim to provide deeper insights into the origins of the depressed dipole mode and the mechanisms behind the formation and evolution of internal stellar magnetic fields.

This paper is structured as follows: Section 2 briefly introduces how to calculate the visibility of dipole modes. Section 3 describes the TESS data and the corresponding data processing methods. Section 4 presents the stars of suppressed dipole mode oscillations found in TESS data and analyzes the results. Section 5 discusses the implications of these findings in the context of stellar magnetic fields and their evolution.

\section{Dipole mode visibilities} \label{sec:vis}

\subsection{General concept}
The mode visibility is usually defined by the ratio of the intrinsic amplitude of radial and dipole modes \citep{Fuller2015Science}:
\begin{equation}
    v_{n_\mathrm{p}, l} = \frac{A_{n_\mathrm{p},l}^2}{A_{n_\mathrm{p},0}^2}.
\end{equation}
Here, $n_\mathrm{p}$ labels a given pressure radial eigenvalue and $l$ stands for the multipole modes. However, for giant stars considered in this paper, mixed mode may spread the energy of a single pressure mode into a wide frequency range. As a result, the amplitude should be considered as a function of frequency $A_{n_\mathrm{p}, l}(\nu)$. In this case, if the driving is the same for all the modes, the total visibility of all the mixed modes is the same as the corresponding “pure" pressure mode \citep{Mosser2015AA}:
\begin{equation}
    \label{eq:mix_vis}
    v_{n_\mathrm{p}, l} = \sum_{m} v_{n_\mathrm{p}, l, m},
\end{equation}
where $m$ marks the mixed modes related to $n_\mathrm{p}$ and $l$. 
Using Eq.\ref{eq:mix_vis}, we can integrate across the multipole mode range to obtain the visibility from spectra:
\begin{equation}
    \label{eq:vis_int}
    v_{n_\mathrm{p},l} = \int_{\nu^\mathrm{min}_{n_\mathrm{p},l}}^{\nu^\mathrm{max}_{n_\mathrm{p},l}}
    \frac{A_{n_\mathrm{p},l}^2(\nu)}{A_{n_\mathrm{p},0}^2(\nu)} \mathrm{d}\nu,
\end{equation}
where $\nu^\mathrm{min}_{n_\mathrm{p},l}$ and $\nu^\mathrm{max}_{n_\mathrm{p},l}$ are the lower and upper bound of the mode corresponding to eigenvalue $l$ and $n_\mathrm{p}$. No matter whether one has mixed mode or not, Eq. \ref{eq:vis_int} is valid. In this paper, we will mainly focus on the visibility of the dipole mode (denoted simply by $v_{n_\mathrm{p}}$):
\begin{equation}
    \label{eq:dipole_vis}
    v_{n_\mathrm{p}, 1} = \int_{\nu^\mathrm{min}_{n_\mathrm{p},1}}^{\nu^\mathrm{max}_{n_\mathrm{p},1}}
    \frac{A_{n_\mathrm{p},1}^2(\nu)}{A_{n_\mathrm{p},0}^2(\nu)}\mathrm{d}\nu.
\end{equation}

\subsection{Depressed Dipole Mode}
The depressed dipole mode is generally attributed to two factors: energy leakage from the envelope region (acoustic cavity) to the core and the suppression of g-mode oscillation in the central part. While the dynamic of g mode suppression remains unclear, the energy transmission due to the mixed mode in giant stars has been thoroughly discussed \citep{Mosser2016AAp, Fuller2015Science}. Here, we breifly review the results.

When the oscillation in the core is fully suppressed, under the Wentzel-Kramers-Brillouin (WKB) approximation, the transmission coefficient $T$ is:
\begin{equation}
        T = \mathrm{exp}\left[ \int_{r_\mathrm{1}}^{r_\mathrm{2}} i k_\mathrm{r} \mathrm{d} r \right].
\end{equation}
Here, the value of transmission coefficient $T^2$ stands for the fractional decrease in oscillation energy when the wave travels across the evanescent region and $k_\mathrm{r}$ (an imaginary number) is the radial wavenumber within the evanescent region. Also, $r_{1,2}$ are the lower and upper bounds of the evanescent region.

Now we know that the energy leakage would cause the oscillation modes in the acoustic cavity to lose a fraction $T^2$ of its energy in a time of 2$t_\mathrm{cross}$, where $t_\mathrm{cross}$ is the crossing time in the acoustic cavity. So this can be equally considered as a damping rate of $\frac{T^2}{2t_\mathrm{cross}}$. At the same time, acoustic oscillations also have a canonical damping rate $\gamma_\mathrm{ac}$. As the amplitude would be inversely proportional to the total damping rate of the corresponding mode, assuming all the modes have the same $\gamma_\mathrm{ac}$, we can express the dipole mode visibility as:
\begin{equation}
    \begin{aligned}
    v_{n_\mathrm{p}} 
    &= \frac{\gamma_\mathrm{ac}}{\gamma_\mathrm{ac}+\frac{T^2}{2t_\mathrm{cross}}}\\
    &= \frac{1}{1+\tau\Delta\nu\, T^2}, \label{eq:dnuT2}
\end{aligned}
\end{equation}

where $\tau = 1/\gamma_\mathrm{ac}$ is the canonical damping time in the acoustic cavity and we use the relation $\Delta \nu = 1/2t_\mathrm{cross}$ between crossing time and the large frequency separation. 

While quadrupole and higher multipole modes can also be suppressed, the effect is much weaker because the transmission factor $T^2$ is smaller for those modes. Due to the large noise level, suppression of quadrupole and higher multipole modes is difficult to observe in TESS data.

\section{Target Selection and Method} \label{sec:method}

\begin{figure*}[htbp]
\centering
\includegraphics[width=0.9\textwidth]{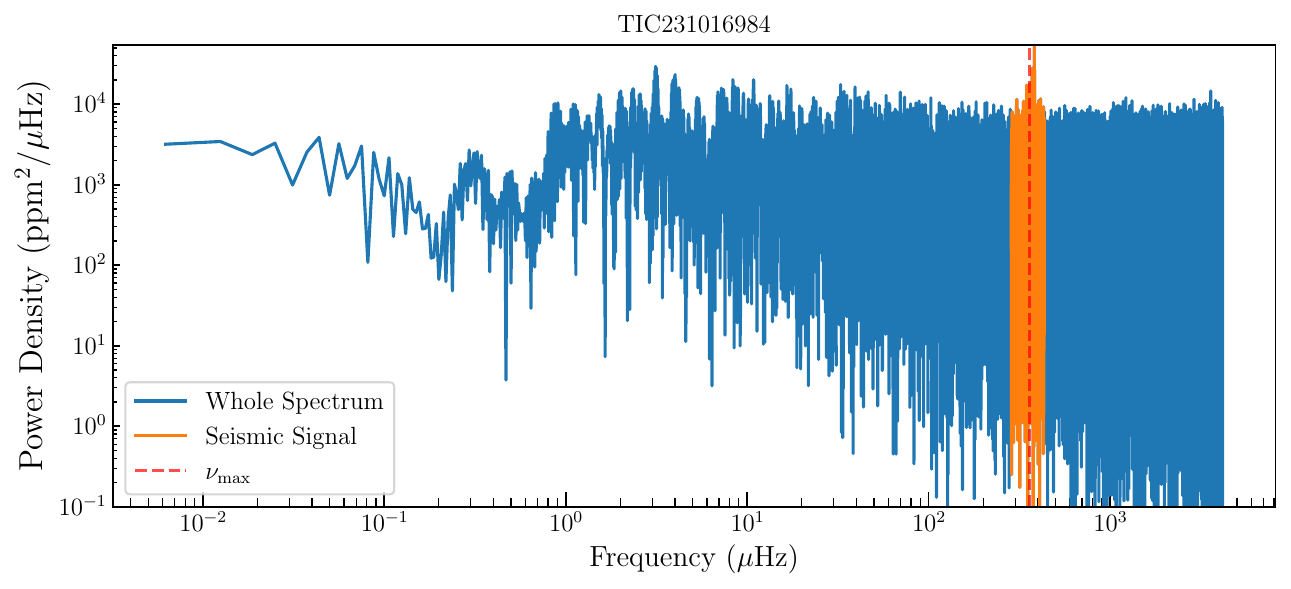}
\includegraphics[width=0.9\textwidth]{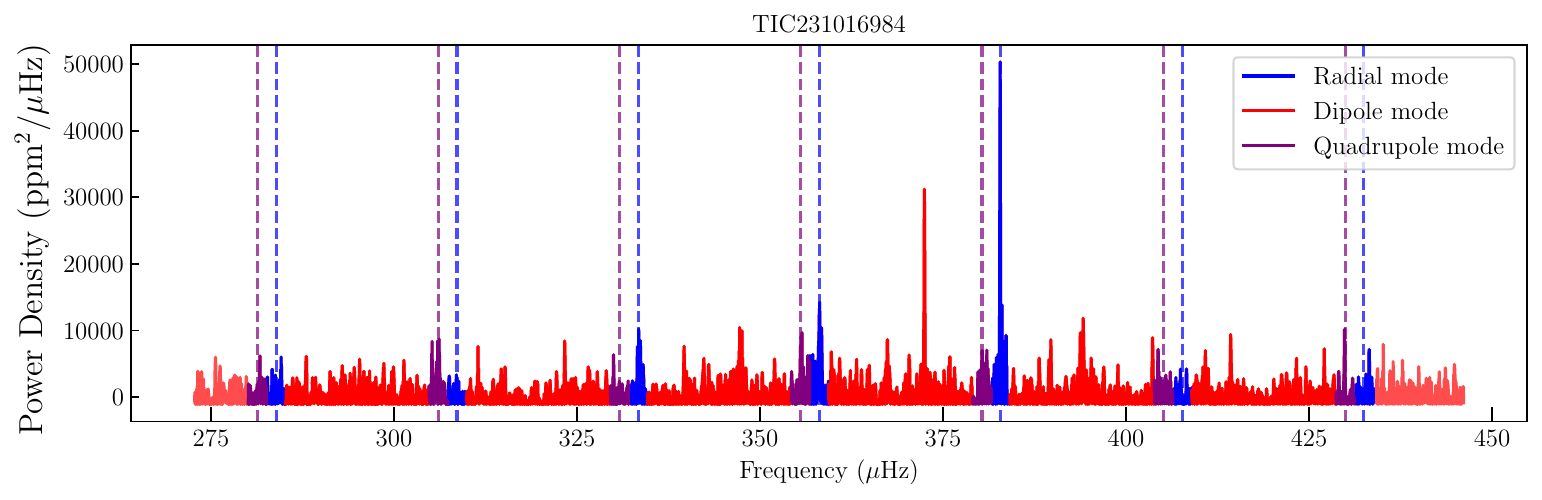}
\includegraphics[width=0.9\textwidth]{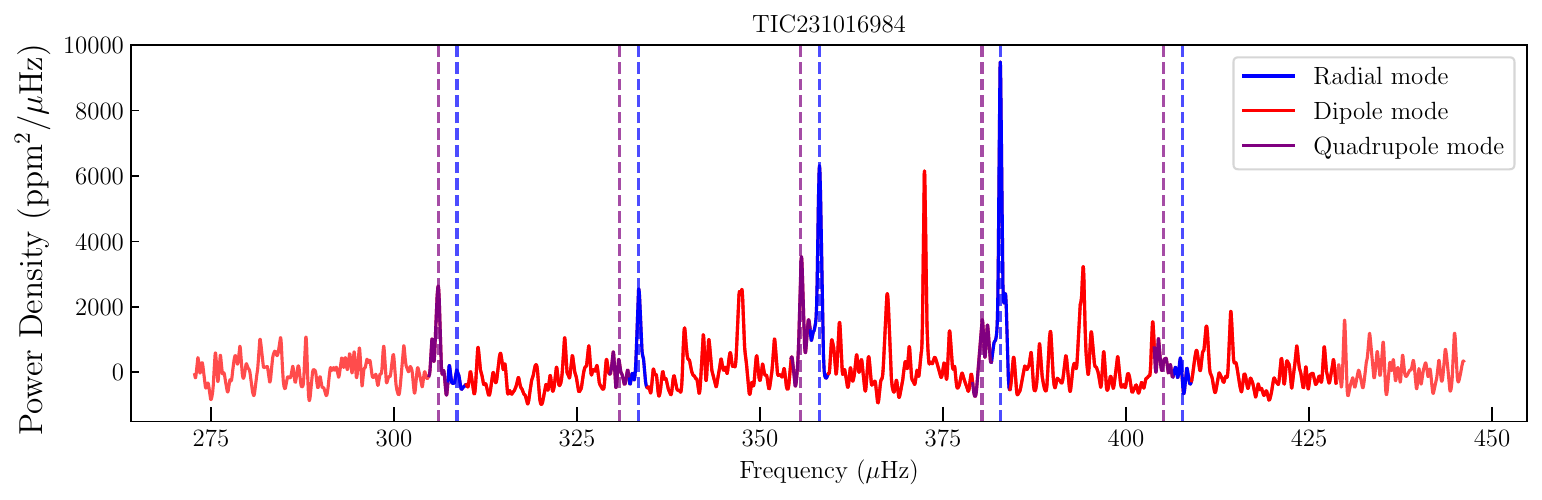}
\caption{Power density spectrum analysis of TIC 231016984.
The top panel shows the full power density spectrum in blue, with the seismic signal highlighted in orange, and the \numax{} marked by the dashed red line. 
The middle panel presents the zoomed-in region of the seismic signal, showing the mode identification results with the background-subtracted spectrum (but not smoothed). Here, blue, red, and purple indicate radial, dipole ($l=1$ region we defined in the text), and quadrupole modes, respectively. The light red area at the beginning and the end of the spectrum indicates the range outside of our mode identification region. Also, we use dashed lines to show the central frequency of radial and quadrupole modes.
The bottom panel displays the modes in the smoothed spectrum (with background subtracted) used to calculate the visibility of the oscillation modes, with the same mode classifications as in the middle panel.}
\label{fig:spec_process}
\end{figure*}

\subsection{TESS Targets} \label{subsec:TESS}

We study the solar-like oscillators identified by \cite{Zhou2024APJ}, who found 8,651 evolved stars with significant detection. \citet{Zhou2024APJ} also provided  the frequency at maximum power (\numax{}) for all stars and valid large frequency separation (\Dnu{}) for 7,509 of them. We downloaded all available short-cadence Pre-search Data Conditioning Simple Aperture Photometry (PDCSAP) flux light curves for these stars (including 2 minutes data with a Nyquist frequency of  4167 \muhz{}) and 20-second data with a Nyquist frequency of  25000 \muhz{}) released before March 2024, spanning from Sectors 1 to 75, from the
Mikulski Archive for Space Telescopes (MAST)\footnote{Data can be found in MAST: \dataset[10.17909/t9-nmc8-f686]{http://dx.doi.org/10.17909/t9-nmc8-f686}}.
These light curves were processed by the TESS Science
Processing Operations Center (SPOC) pipeline 
\citep{Twicken2016ANJ,Jenkins2020}.
We did not mix the 2-minute and 20-second light curves but analyzed the two types of data independently. At the same time, we only use the stars that have been observed in more than 4 sectors (i.e. $N_\mathrm{sec}>4$). The number of stars in with 2-minute and 20-second light curves in different steps of our processing is summarized in Table \ref{table:yields}. 
\begin{table}[h]
\caption{The table summarizes the number of stars within different kinds of selection criteria across our analysis. 120s (20s) indicates the stars with 120s (20s)-cadence light curves.}
\makebox[0.4\textwidth][c]{
\begin{tabular}{lrrr}
\toprule
Criteria&120s&20s&All\\
\midrule
All stars with $\Delta \nu$  & 7505 & 1674 & 7509 \\
$N_\mathrm{sec}>4$  & 2929 & 316 & 3115 \\
$V_1<0.8$  & 559 & 112 & 591 \\
Significant suppression  & 170 & 26 & 179 \\
Potential suppression  & 113 & 20 & 119 \\
11 high \numax{} stars & 9 & 2 & 11 \\
\bottomrule
\end{tabular} }
\label{table:yields}
\end{table}

\subsection{Fitting Power Spectrum}
\label{subsec:bg}

We use the Lomb-Scargle periodogram method \citep{VanderPlas2018APJS} to transform the light curves into the power spectra. The spectra can be decomposed into three components: the granulation background $B(\nu)$, an oscillation power excess $P_\mathrm{G}(\nu)$, and white noise $W_\mathrm{n}$ \citep{Mathur2011APJ,Chaplin2014APJS}.

For the power excess, we use a Gaussian envelope centered at \numax{}:
\begin{equation}
    P_\mathrm{G}(\nu) = P_\mathrm{g}\mathrm{exp}
    \left(-\frac{(\nu_\mathrm{max}-\nu)^2}{2\sigma_\mathrm{g}^2}\right),
    \label{eq:Pg}
\end{equation}
where $P_\mathrm{g}$ describes the height at \numax{} and $\sigma_\mathrm{g}$ is the standard deviation of the envelop \citep{Michel2008Science}.

As for the convection, we take three Harvey-like functions \citep{Harvey_func}:
\begin{equation}
        B(\nu) = 
    \sum_{i = 1}^{3}\frac{4 \sigma_i^2 \tau_i}{1+(2\pi \nu \tau_i)^4},
\end{equation}
where $\sigma_i$ stands for the amplitude of $i$th background component and $\tau_i$ is the characteristic time scale for the $i$th components.

Following the above description, the decomposition of power spectra is:
\begin{equation}
    \begin{aligned}
        P_\mathrm{model}(\nu) = 
     \eta(\nu)^2
    &
    \left[\sum_{i = 1}^{3}
    \frac{4 \sigma_i^2 \tau_i}
    {1+(2\pi \nu \tau_i)^4}\right.
    \\
    &
    + \left .P_\mathrm{g}\mathrm{exp}\left(-\frac{(\nu_\mathrm{max}-\nu)^2}{2\sigma_\mathrm{g}^2}\right)
    \right ]
    +W_\mathrm{n},
    \end{aligned}
\end{equation}
where $\eta(\nu) = \mathrm{sinc}(\pi \nu/2 \nu_\mathrm{nyq})$ stands for the attenuation due to the observational signal
discretization, and $\nu_\mathrm{nyq}$ is the Nyquist frequency \citep{Kallinger2014AAp}.

We did Bayesian inference with Markov Chain Monte Carlo (MCMC) to accurately fit eight different parameters ($\sigma_{1-3}$, $\tau_{1-3}$, $P_\mathrm{g}$, and $\sigma_\mathrm{g}$, \numax{} is provided by the catalog we use) with Python package \texttt{EMCEE} \citep{emcee}. In order to separate the three Harvey components properly, we fixed $\tau_1 > \tau_2 > \tau_3$ in the whole MCMC process and set the initial value of $\tau_{1-3}$ to be [$0.8 \nu_\mathrm{max}$, $\nu_\mathrm{max}$, $1.2 \nu_\mathrm{max}$]. We choose a Gaussian likelihood for the MCMC simulation:
\begin{equation}
    \begin{aligned}
        \mathrm{log}\mathcal{L} \propto \sum_{j} [P_\mathrm{obs}(\nu_j) - P_\mathrm{model}(\nu_j)]^2,
    \end{aligned}
\end{equation}
where $\nu_j$ is the frequency of the $j$th data point and $P_\mathrm{obs}$ is the spectrum transformed from the TESS light curve. Here, we smoothed $P_\mathrm{obs}$ by a moving box in logarithmic space, to weaken the uncertainty caused by the noise of the original spectrum. Also, as we focus on the seismic signal, we fit the background within [$\nu_\mathrm{max}-5\Delta\nu$, $\nu_\mathrm{max}+5\Delta\nu$].

For each spectrum, we ran 3000 steps MCMC with 1500 burn-in steps. We take the averaged $\sigma_{1-3}$ and $\tau_{1-3}$ from the MCMC and subtracted the Harvey background:
\begin{equation}
    P_\mathrm{sub} = P_\mathrm{obs} - 
    \sum_{i = 1}^{3} \frac{4 \bar{\sigma_i}^2 \bar{\tau}_i} {1+(2\pi \nu \bar{\tau}_i)^4}.
\end{equation}

\begin{figure*}[htbp]
\centering
\includegraphics[width=0.9\textwidth]{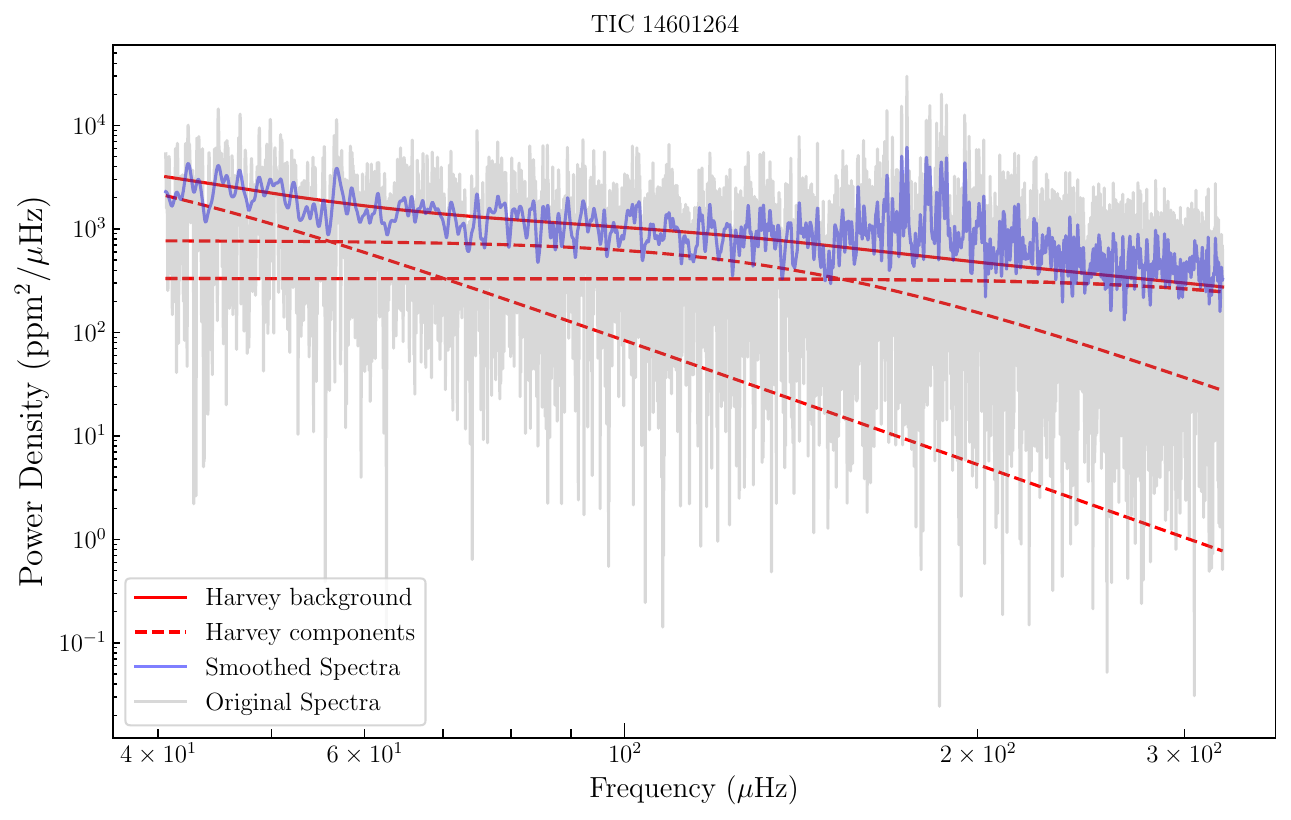}
\caption{Background fitting for TIC 14601264, showing the original power spectrum (gray), smoothed power spectrum (blue), and the fitted Harvey function components (red dashed lines) along with the combined background model(solid red line). The frequency range is zoomed to show the fitting result near \numax{}.
}
\label{fig:bg_fitting}
\end{figure*}

An example of the background spectrum and the range of power excess is shown in the top panel of Fig.\ref{fig:spec_process}, where the whole spectra of the star TIC 231016984 are shown in blue and the seismic signal is marked with orange. We also show the \numax{} from \citep{Zhou2024APJ} by the red dashed line. There is a slight offset between the \numax{} and the actual peak of the power spectrum, as \cite{Zhou2024APJ} used the central frequency of the fitted Gaussian envelope to represent \numax{}. However, due to mode suppression, achieving a perfect fit was more challenging. 
Here, we present the background fitting result for TIC 14601264 in Fig.~\ref{fig:bg_fitting}, where the original spectrum is displayed in gray, the Gaussian-smoothed spectrum in blue, and the individual and combined Harvey functions in red dashed and solid lines, respectively. As can be seen, the inclusion of three Harvey components enables our fitting to effectively reconstruct the background near $\nu_\mathrm{max}$, ensuring sufficiently accurate background subtraction for subsequent analysis.

\subsection{Asteroseismic Mode Identification} \label{subsec:pbjam}

We employ the open-sourced Python package \texttt{PBjam}\footnote{\dataset[https://pbjam.readthedocs.io/en/latest/index.html]{https://pbjam.readthedocs.io/en/latest/index.html}} to identify and fit the solar-like oscillation modes on the power spectrum \citep{PBjam}. 
\texttt{PBjam} takes only three global seismic properties as input: the frequency of maximum power (\numax{}), large frequency separation \Dnu{}, and the effective temperature of the star ($T_\mathrm{eff}$), or the Gaia photometric color index ($G_\mathrm{BP}-G
_\mathrm{RP}$). As $T_\mathrm{eff}$ and $G_\mathrm{BP}-G_\mathrm{RP}$ contain similar information, only one of them is necessary.
 With the above inputs, \texttt{PBjam} would output the fitting value of \numax{} and \Dnu{} together with the following parameters: the radial order phase term ($\varepsilon$), the frequency difference of ($l = 0$) and ($l = 2$) modes ($\delta\nu_{02}$), scale (curvature) of radial order variation with frequency ($\alpha$), the height of the Gaussian envelope ($H_{\mathrm{max}}$), the width in the frequency of the envelope ($W_{\mathrm{env}}$) and the width of the oscillation modes ($\delta \nu$). Aside from above parameters, it would also output a list of the central frequencies of radial and quadruple modes. Detailed explanations of the fitting process are available in \cite{PBjam}.

In our fitting process, we feed the original power spectrum (without background subtraction), together with the \numax{}, \Dnu{}, and the $T_\mathrm{eff}$ from the catalog of \citet{Zhou2024APJ} into \texttt{PBjam}, and then take 7 pairs of the frequencies of radial and quadrupole modes around \numax{}  ($\nu_{n_\mathrm{p},0}$ and $\nu_{n_\mathrm{p},2}$ respectively) from the outputs.  
We show an example of the mode identification result in the middle panel of Fig. \ref{fig:spec_process}, where seven pairs of modes centred at the \numax{} are shown. As shown in the figure, we take the range $[\nu_{n_\mathrm{p},0}-3\delta \nu, \nu_{n_\mathrm{p},0}+3\delta \nu]$ to be the radial mode range and $[\nu_{n_\mathrm{p},2}-3\delta \nu, \nu_{n_\mathrm{p},2}+3\delta \nu]$ for the quadrupole mode.
Furthermore, we consider the dipole mode distribution across the interval between the radial and quadrupole modes, specifically within the range $[\nu_{n_\mathrm{p},0}+3\delta \nu, \nu_{n_\mathrm{p}+1,2}-3\delta \nu]$. This approach ensures comprehensive coverage of the dipole modes. As for the $l=3$ modes inside our dipole region, due to the $\sim10^{-2}$ visibility of $l=3$ modes \citep{Ballot2011AAp}, their inclusion would not affect our results significantly.

\subsection{Dipole Mode Visibility}
Now, we proceed to work with background subtracted spectra. Due to the noise level in the original spectra, we applied a Gaussian kernel with the width of $\Delta \nu$ to smooth the spectra. Based on the mode identification for $l = 0, 2$, and the $l=1$ region between them, we calculate the visibility.
Due to the limited observing time, the excitation and damping of a single mode may not be fully resolved. In this case, we take the average of three modes around \numax{} with the smoothed spectrum:
\begin{equation}
        V_1 =  \frac{\sum_{i=-1}^{1} v_{n_i,1}}{\sum_{i=-1}^{1} P_\mathrm{G}(\nu_i)} .
\end{equation}

Here, $v_{n_i,1}$ is given by the integral of Eq. (\ref{eq:dipole_vis}), within the radial, dipole, and quadrupole range define in Sec \ref{subsec:pbjam},
and the denominator is chosen to eliminate the effect of the Gaussian envelope. 
$P_\mathrm{G}$ is the same in Eq.\ref{eq:Pg} and $\nu_i$ is the central  frequency of each dipole region.
We show the modes that we choose to calculate visibility for star TIC 231016984 in the bottom panel of Fig. \ref{fig:spec_process}.

Previous observations and simulations of Kepler data have shown that the visibility of dipole modes without suppression is typically around 1.5, though it can vary based on the effective temperature of the star \citep{Ballot2011AAp}. Depressed dipole modes, on the other hand, tend to have visibility values around 0.5 for \numax{} =70 $\mu\mathrm{Hz}$ and would decrease with \numax{} \citep{Stello2016Nature}. However, due to the substantial noise in TESS data and the incomplete mode excitation spectrum caused by the relatively short observation periods, obtaining a precise estimate of visibility is challenging. Therefore, relying solely on visibility as a threshold for identifying depressed dipole modes is not the most reliable approach for TESS data.

\begin{figure}[htbp]
\centering
\includegraphics[width=0.48\textwidth]{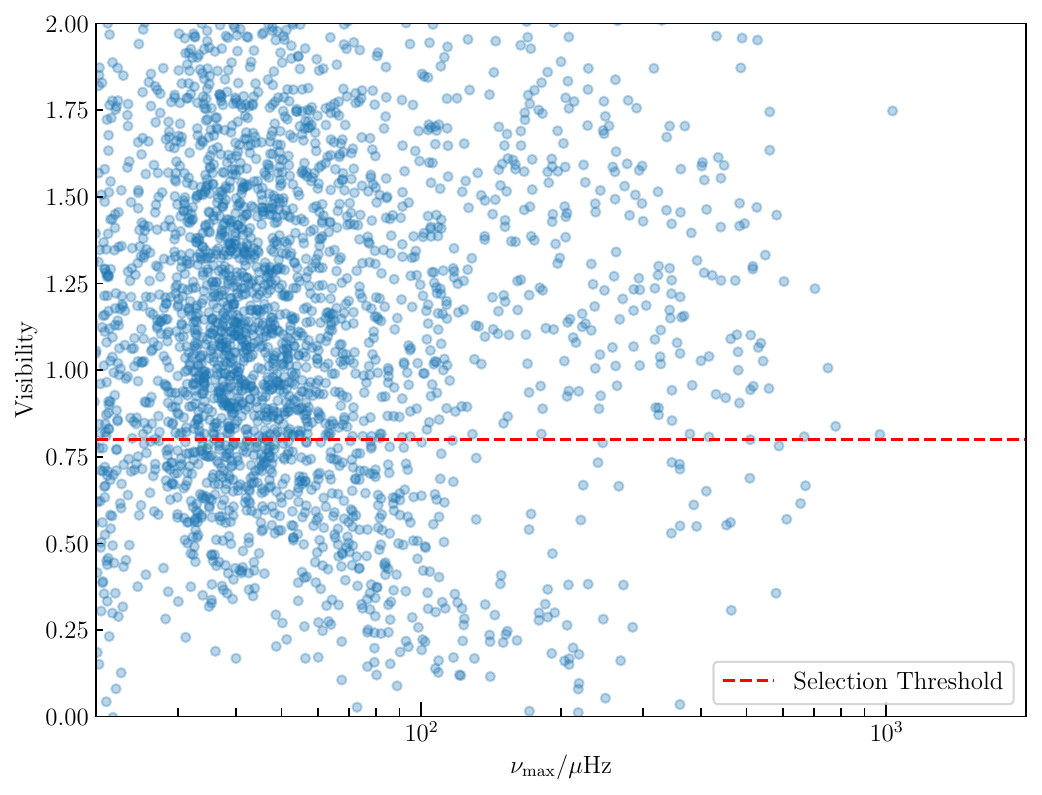}
\includegraphics[width=0.48\textwidth]{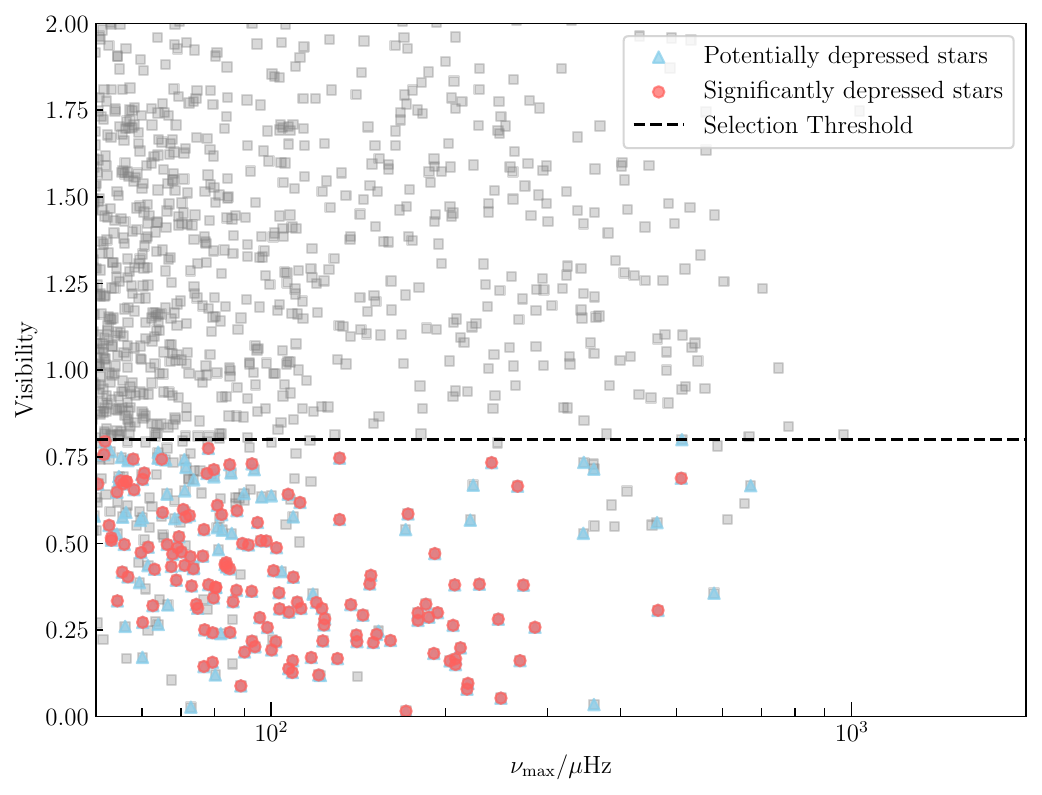}
\caption{The upper panel shows the visibility distribution of all stars in the sample. The red dashed line represents the selection threshold for depressed dipole modes, set at \( V < 0.8 \). In the lower panel, we highlight stars with significantly depressed dipole mode (red) and potentially depressed mode (blue), alongside all seismic targets with measurable visibility (grey), with the \numax{} range restricted to \numax{} $> 50 \, \mu\mathrm{Hz}$. The black dashed line again indicates the selection threshold. Due to the low signal-to-noise ratio (SNR) in the TESS data and the short observation period, we are unable to observe a clear two-band structure in this sample. But as shown in the bottom panel, most depressed stars with high \numax{} have visibility smaller than $0.5$, indicating $V<0.8$ is a rather loose threshold.}
\label{fig:vis}
\end{figure}

To ensure we capture all stars with potential dipole mode suppression, we initially applied a rough visibility threshold of 0.8 to identify candidates. This selection was then refined through manual inspection to improve accuracy. During the manual inspection, spectra exhibiting clear Gaussian envelopes and significantly reduced dipole mode amplitudes were classified as having significant suppression. In contrast, spectra with irregular envelope shapes or missing radial modes were either discarded as false positives or categorized as potentially suppressed cases requiring further investigation.
Notice that this step only gets rid of 29.3\% of our low visibility sample stars about \numax{} $>$ 50 $\mu\mathrm{Hz}$ (see Fig. \ref{fig:vis}), and so it has no significant impact on our main results.

\subsection{Asteroseismic Modelling}
\label{subsec:MESA}
We use an established stellar model grid to constrain the fundamental parameters of stars exhibiting dipole mode suppression. The model grid was originally calculated by \cite{Li2023MNRAS} and included five varying input parameters: mass (\( M/M_{\odot} \)), initial helium fraction (\( Y_\mathrm{init} \)), metallicity ([Fe/H]), the mixing-length parameter (\( \alpha_{\rm MLT} \)), and the overshooting parameter (\( f_{\rm ov} \)). The original grid was designed to test model effects, and thus the parameter ranges are relatively broad. For this study, we only use a subset of the grid, restricting the parameters to the following ranges: \( M/M_{\odot} \sim [0.7, 2.4] \), \( Y_\mathrm{init} \sim [0.25, 0.32] \), [Fe/H] \( \sim [-1.0, 0.5] \), \( \alpha_{\rm MLT} \sim [1.7, 2.5] \), and \( f_{\rm ov} \sim [0.0, 0.02] \). Theoretical modes for \( l = 0 \) and \( l = 2 \) are computed for each grid point.

We adopted global parameters such as effective temperature (\( T_{\mathrm{eff}} \)), [Fe/H], large frequency separation (\( \Delta\nu \)), frequency of maximum power (\( \nu_\mathrm{max} \)), and small frequency separation (\( \delta\nu \)), along with individual mode frequencies for \( l = 0 \) and \( l = 2 \) extracted using \textsc{PBjam}, as input constraints in our fitting scheme. The fitting method follows \cite{Li2017MNRAS}, using maximum likelihood estimation while accounting for systematic uncertainties in individual mode frequencies.

We modeled a total of 591 stars, all below the threshold of $V_1<0.8$. However, we excluded the fitting results for two cases: first, when a star had no satisfactory fit (i.e., the maximum likelihood of individual modes was smaller than \( 10^{-6} \)); second, when the mass converged below 0.8 or above 2.3 \( M_{\odot} \) to avoid edge effects. It is important to note that the modeling results for stars with relatively high luminosity are not reliable, as they lie in the mixing region of H-shell burning and He-core burning stars on the Hertzsprung–Russell diagram. In our fits, all stars are assumed to be H-shell burning. 

In the end, we obtained model-based stellar parameters (mass, age, radius, and surface gravity) for 452 stars, which are listed in Table \ref{table:stars}. Some stars were omitted due to missing \( T_{\mathrm{eff}} \) values or because they reached the edge of the grid. The typical uncertainties are approximately 8\% for mass, 30\% for age, 3\% for radius, and 0.1 dex for surface gravity.
\begin{table}[h]
\centering
\caption{Global seismic properties from asteroseismic modeling. Only the first 10 stars are shown here. The full table can be found online (in the published version of APJ).}

\begin{tabular}{lrrrrr}
\toprule
TIC & \numax{} & \Dnu{} & $T_\mathrm{eff}$ & $M$ & log$(g)$ \\
 & ($\mu$Hz) & ($\mu$Hz) & (K) & ($M_\odot$) & (dex) \\
\midrule
456868486 & 153.2 & 11.5 & -- & -- & -- \\
458484258 & 94.9 & 8.2 & 4705 & 1.0 & 2.8 \\
459678581 & 103.3 & 8.7 & -- & 1.5 & 2.9 \\
459893929 & 29.9 & 3.8 & 4980 & 0.9 & 2.4 \\
459978312 & 101.0 & 7.9 & 4769 & -- & -- \\
459984696 & 61.4 & 5.8 & -- & 1.6 & 2.7 \\
460680010 & 111.0 & 8.8 & -- & 1.3 & 2.9 \\
461636387 & 37.7 & 4.2 & 4599 & 1.1 & 2.5 \\
461652875 & 48.3 & 4.9 & -- & 1.4 & 2.6 \\
461837097 & 249.1 & 19.6 & 5004 & 1.0 & 3.3 \\
\bottomrule
\end{tabular}

\label{table:stars}
\end{table}

\begin{figure*}[htbp]
\centering
\includegraphics[width=0.9\textwidth]{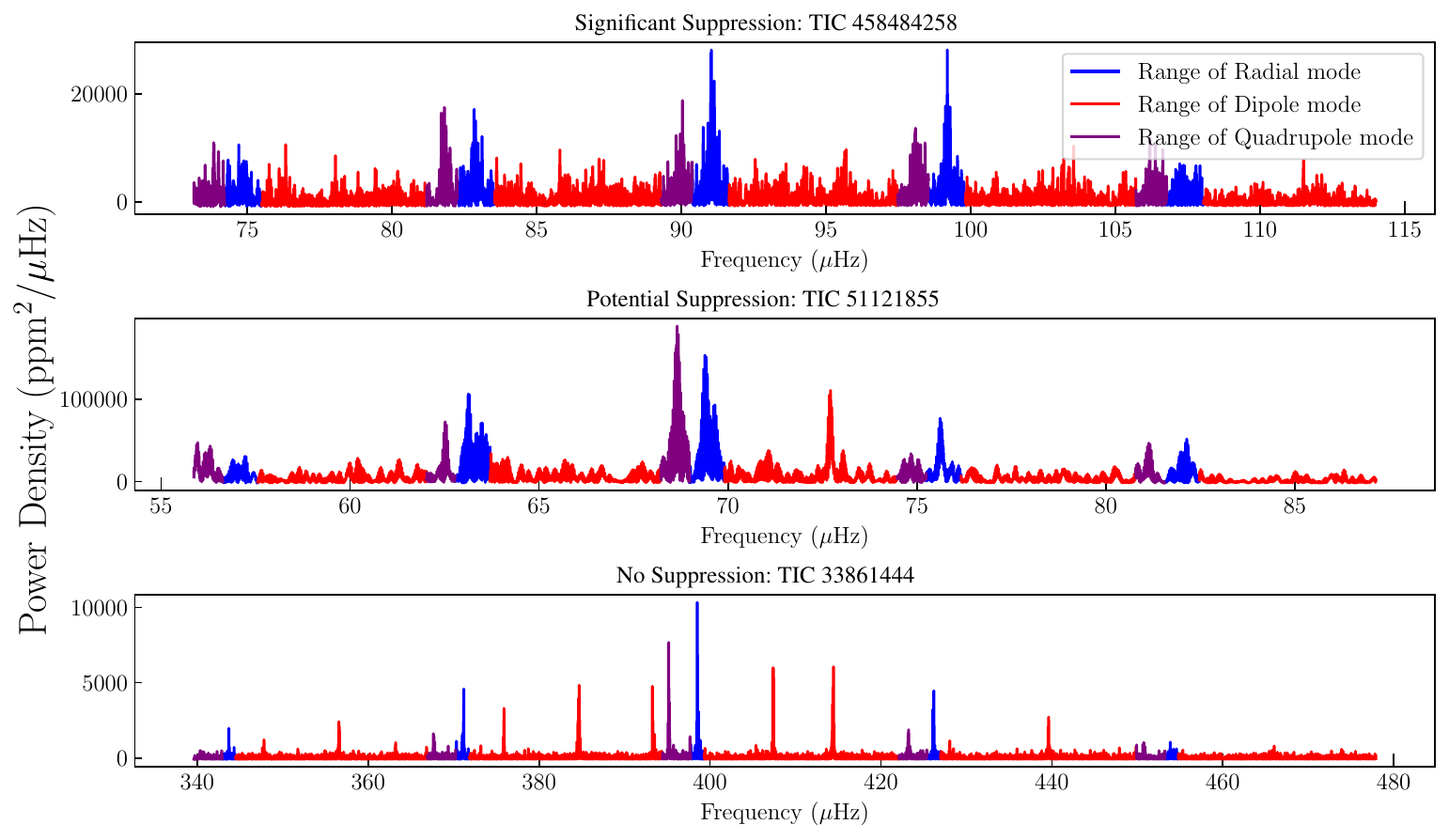}
\caption{Oscillation spectra for different star categories by rank. The spectra are color-coded to distinguish between the range of radial (blue), dipole (red), and quadrupole (purple) modes. The three panels represent stars with different levels of suppression: significant suppression, potential but incomplete suppression, and no suppression. 
}
\label{fig:rank_eg}
\end{figure*}

\begin{figure*}[htbp]
\centering
\includegraphics[width=0.7\textwidth]{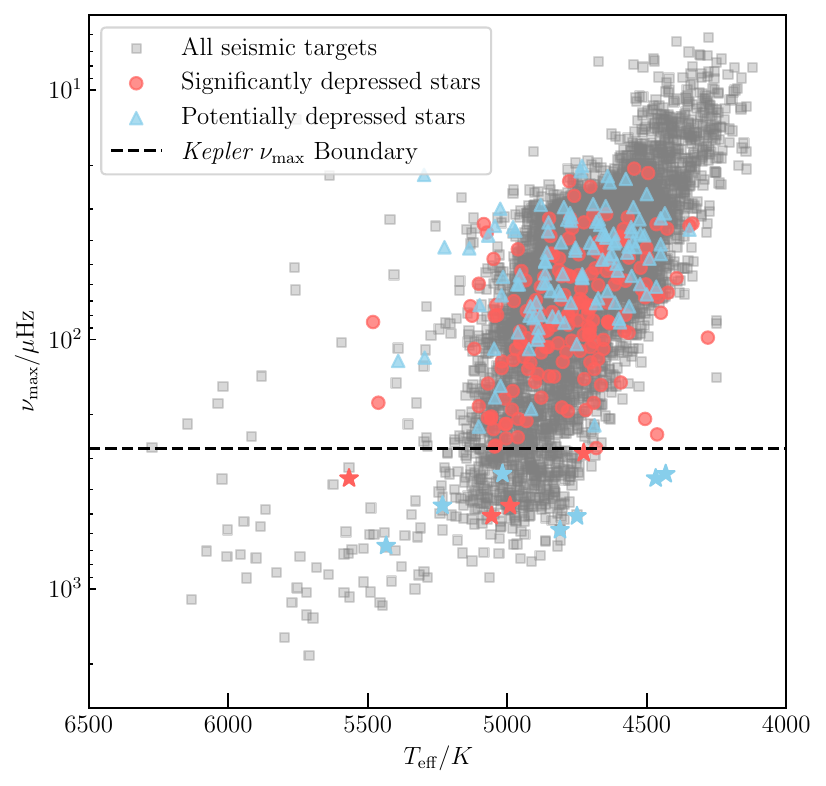}
\caption{TESS Solar-like oscillators on the \teff -\numax{} diagram. The diagram categorizes stars into two ranks based on the suppression of dipole modes: Significantly depressed stars (red) and potentially depressed stars due to short observation time (blue). The grey squares stand for the remaining stars from the TESS catalog. The black dashed line indicates the highest \(\nu_{\mathrm{max}}\) observed in Kepler long cadence data (\numax{= 273.16 $\mu$Hz}), highlighting 11 early-stage red giants and subgiants with significant suppression at frequencies up to 670 $\mu$Hz (highlighted with stars symbols).}
\label{fig:HR_diagram}
\end{figure*}

\begin{figure}[htbp]
\centering
\includegraphics[width=0.48\textwidth]{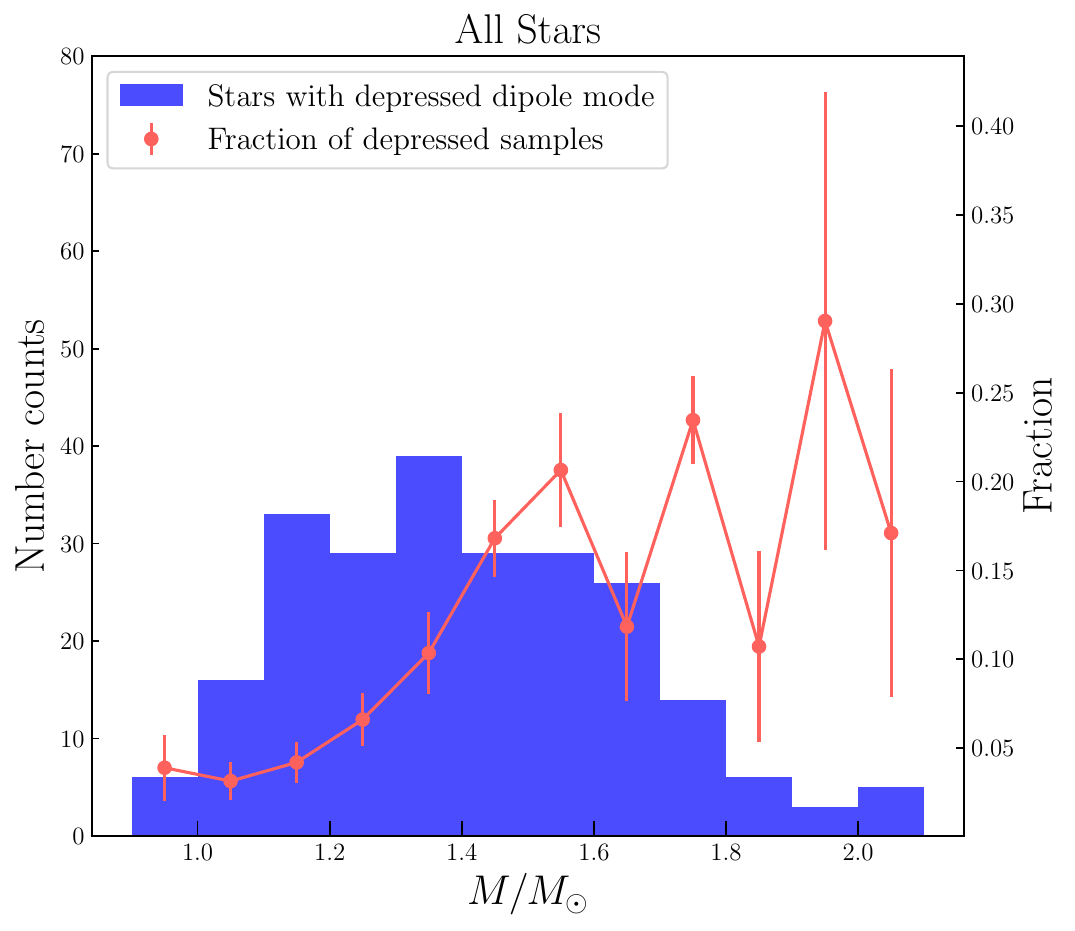}
\includegraphics[width=0.48\textwidth]{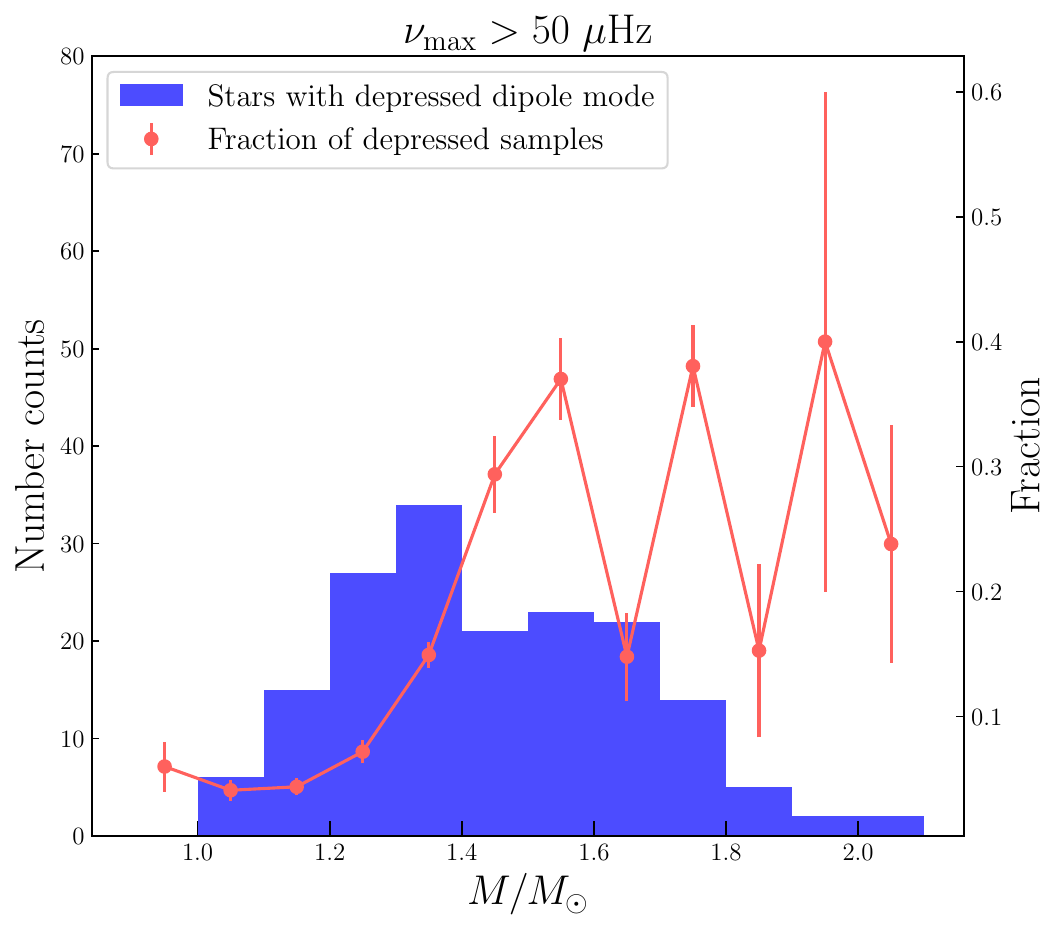}
\caption{Mass distribution of stars with depressed dipole modes (blue histogram, based on asteroseismic modeling as described in Sec. \ref{subsec:MESA}), and the fraction of such stars relative to all stars with measurable seismic signals (red line, using masses derived from the scaling relation). The histograms represent the number counts in each mass bin, emphasizing the relative frequency of stars with dipole mode suppression across different mass ranges. The upper panel displays the results for the entire sample, while the lower panel focuses on stars with \numax{}$>$ 50 $\mu\mathrm{Hz}$. The red lines illustrate that dipole mode suppression predominantly occurs in stars with masses exceeding 1.2 $M_\odot$.}
\label{fig:mass_dis}
\end{figure}

\section{Results} \label{sec:result}

\subsection{Overview}\label{subsec:overview}

We analyzed a sample of 8,649 evolved stars from the TESS catalog.
We selected the stars with a signal-to-noise ratio (SNR) larger than 5 
(we define SNR as the maximum SNR within [$\nu_\mathrm{max}-3\Delta \nu$, $\nu_\mathrm{max}+3\Delta \nu$]),
together with the $N_\mathrm{sec}>4$ we applied in star selection, as failure in either of these conditions would cause a noisy spectrum. We successfully identified individual modes for 2,633 stars. The result of visibility is shown in Fig. \ref{fig:vis}. We calculated the visibility for those stars and manually reviewed all the stars under the threshold of $V_1 <0.8$. As mentioned in Sec.\ref{subsec:MESA}, there are 591 stars below the threshold. After manual the selection based on whether the spectra have non-Gaussian envelope and unresolved modes, we found 179 stars with significant dipole mode depression. We show significantly depressed stars with \numax{}$>50$ in the bottom panel of Fig. \ref{fig:vis} with purple. Though there is a huge scatter in the overall visibility distribution, most of the depressed stars with large \numax{} have visibility below $0.5$, indicating our $V<0.8$ threshold can capture basically all the early stage stars with depressed dipole mode. 
Additionally, we found 119 stars with poorly resolved modes showing depressed modes in one or two $l=1$ regions. Those potential dipole mode suppressions are likely due to the limited observation time.
Some examples are demonstrated in Fig \ref{fig:rank_eg}. 

Fig.\ref{fig:HR_diagram} illustrates the distribution of these stars in the \teff{} - \numax{} diagram.
Notably, the dashed line in Fig. \ref{fig:HR_diagram} marks the highest \numax{} ($273.16$ $\mu \mathrm{Hz}$) observed in Kepler long cadence data, highlighting 11 early-stage red giants and subgiants with suppressed dipole modes.
The least-evolved star (TIC 260823248) has a \numax{} of 670 $\mu$Hz. These findings infer that the dipole mode suppression can appear at late-subgiant and early red-giant phases. Also, it might be interesting to use some Super-Nyquist analyses \citep{Murphy2013MNRAS, Chaplin2014MNRAS, Yu2016MNRAS} on Kepler data to look for similar cases.

For the red line, we also present the fraction of stars with depressed dipole modes among all the stars with measurable seismic signals (the number of sectors larger than 4 and SNR larger than 5) as a function of stellar mass in Fig. \ref{fig:mass_dis}. The error bars account for the uncertainty due to the inclusion of potential, but unconfirmed, depressed dipole mode star. The upper limit of the error bars assumes that all potential stars are valid, whereas the lower limit assumes that none are valid.
As we do not have modeling mass for all the stars in the TESS catalog, masses were determined using the scaling relation described by \citet{Zhou2024APJ}:

\begin{equation}
    \frac{\nu_\mathrm{max}}{\nu_{\mathrm{max},\odot}} = \frac{g_\mathrm{fit}}{g_\odot}\left( \frac{T_\mathrm{eff}}{T_{\mathrm{eff},\odot}} \right)^{-0.397} (10^{[\mathrm{M/H}]})^{0.008},
\end{equation}
where \(g_\mathrm{fit}\) represents the surface gravity of the star and $[\mathrm{M/H}]$ is the metallicity. The comparison reveals that dipole mode suppression is more prevalent among stars with higher masses, particularly those above 1.2 $M_\odot$.
 
As seen with Kepler observations, a clear trend for the stars with \numax{}$>$ $50$ $\mu\mathrm{Hz}$ is found: the fraction of stars exhibiting depressed dipole modes increases with mass, reaching a high level after 1.4 $M_\odot$, which is consistent with the result based on the \kepler{} data \citep{Stello2016Nature}.

\subsection{Subgiants and Young Red Giants}\label{subsec:early}

\begin{figure*}[htbp]
\centering
\includegraphics[width=0.9\textwidth]{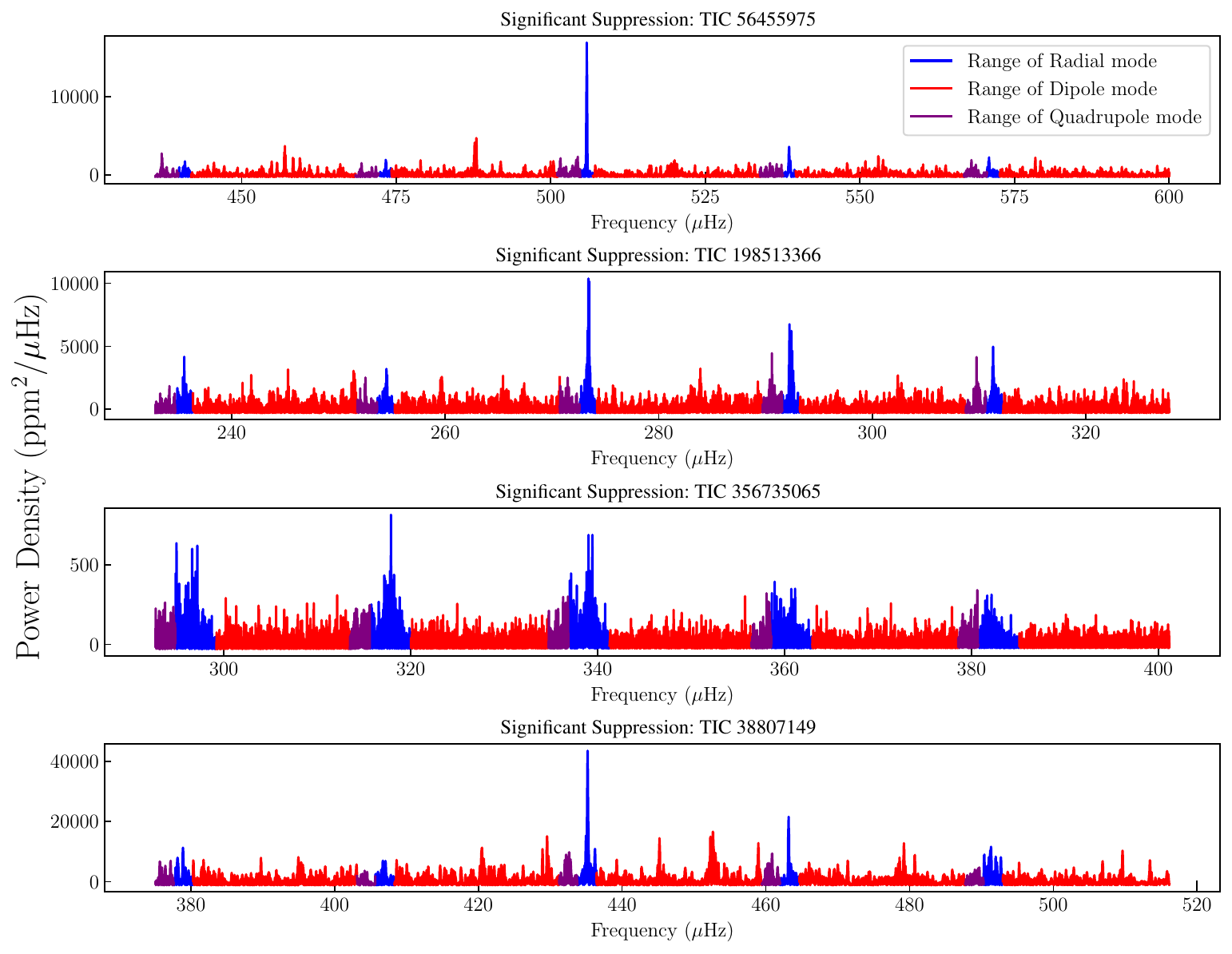}
\caption{Power spectra of 4 stars (TIC 56455975, TIC 198513366, TIC356735065, TIC 38807149) with significant dipole mode suppression, highlighting this phenomenon in early-stage giant stars and subgiants. The spectra are color-coded to distinguish between the range of radial (blue), dipole (red), and quadrupole (purple) modes.}
\label{fig:4_sig}
\end{figure*}

\begin{figure*}[htbp]
\centering
\includegraphics[width=0.9\textwidth]{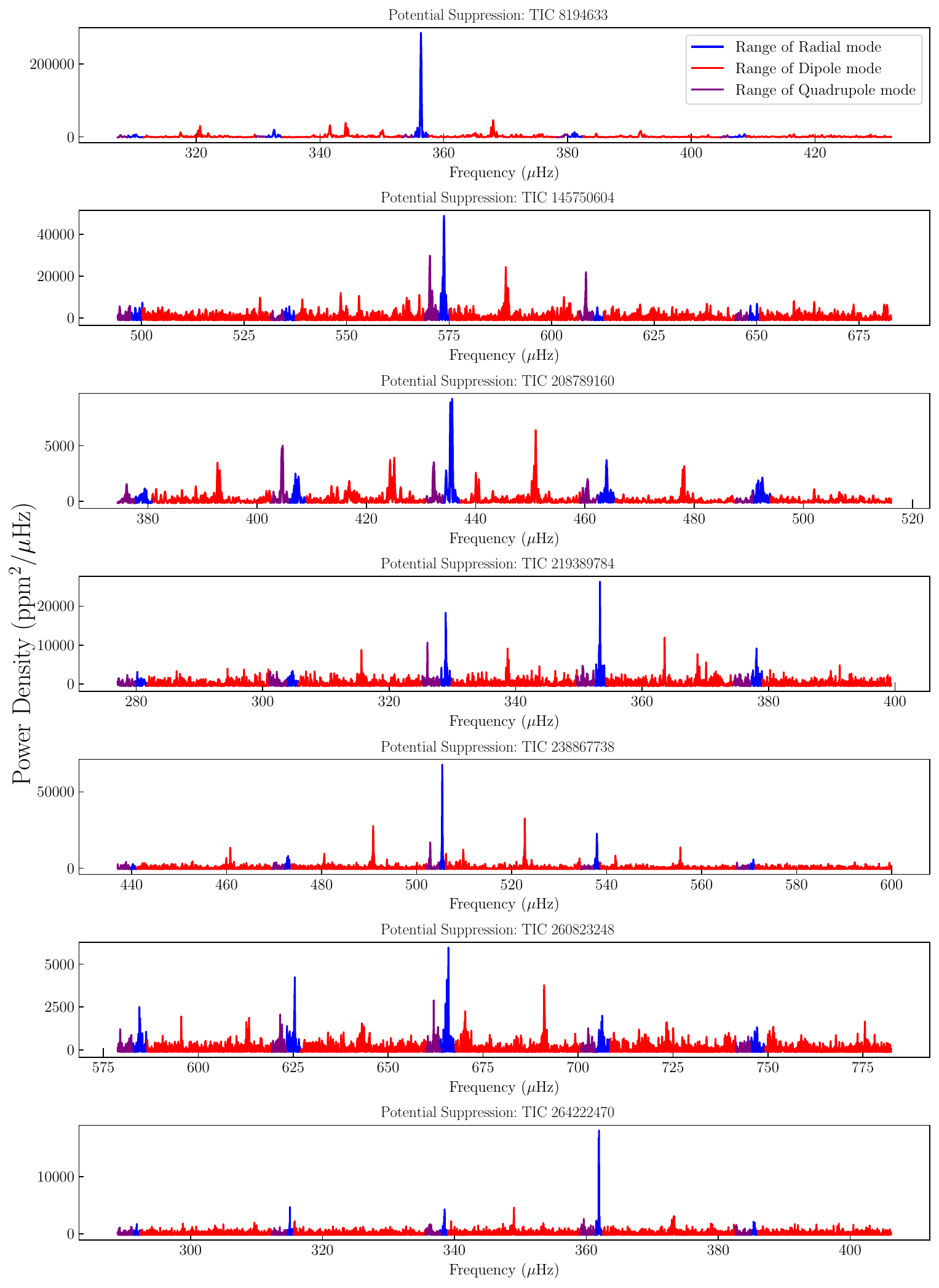}
\caption{Power spectra of 7 stars with potential suppression, whose frequencies are higher than the Kepler observation threshold ($273.16$ $\mu \mathrm{Hz}$). These spectra are color-coded to distinguish between the range of radial (blue), dipole (red), and quadrupole (purple) modes.}
\label{fig:7_pot}
\end{figure*}

Here we discuss the 11 stars with \numax{} $>273.16 $ $\mu\mathrm{Hz}$, up to a maximum \numax{} of 670 $\mu \mathrm{Hz}$ (TIC 260823248). In Fig.\ref{fig:4_sig} and Fig.\ref{fig:7_pot}, we illustrate the power spectra of these 11 stars. As can be seen, four of them (TIC 56455975, TIC 198513366, TIC 356735065, and TIC 38807149, shown in Fig. \ref{fig:4_sig}) exhibit significant dipole mode suppression, which is rarely observed in these early-stage giant stars and subgiants. The presence of this suppression at such an early evolutionary stage is particularly intriguing, as it open a new window for understanding of stellar oscillations and interior structure. We will briefly return to the implications of this result in Sec. \ref{sec:conclusion}.

\begin{figure*}[htbp]
\centering
\includegraphics[width=0.9\textwidth]{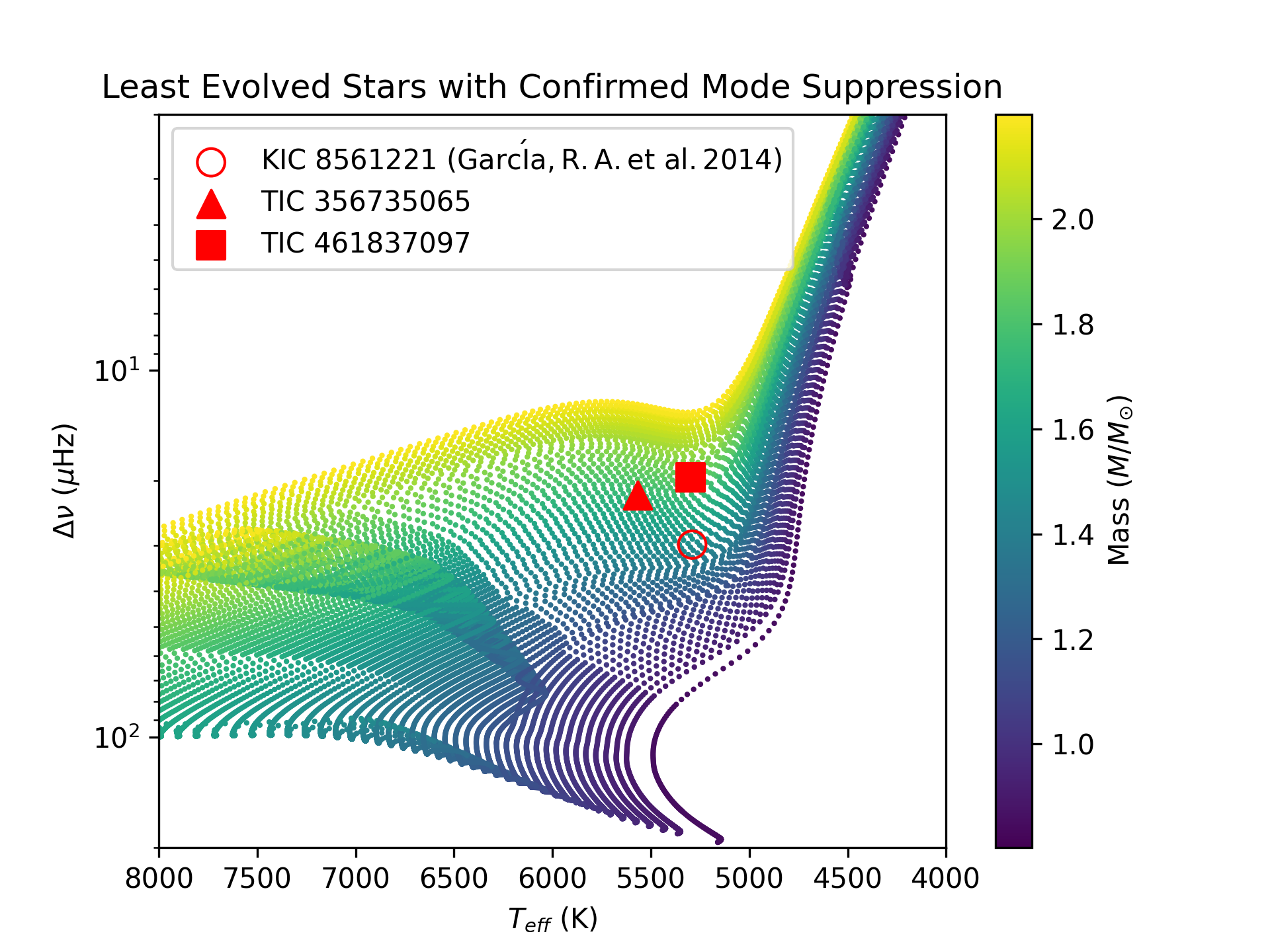}
\caption{Three subgiants with significant dipole-mode suppression in the \teff -\Dnu{} plane. The filled triangle and square refer to the two stars (TIC 356735065 and TIC 461837097) found in this work. The open circle represents the star KIC 8561221 studied by \citet{Garcia2014AAp}. The dotted lines represent the evolution tracks with different masses from 0.8 to 2.2 $M_{\odot}$ (indicated by the color bar).}
\label{fig:least}
\end{figure*}

At the same time, it is also worth mentioning that, the sample TIC 356735065 and TIC 461837097, are likely two late subgiants. The effective temperature and \numax{} of TIC 356735065 are 5567 K and 360 $\mu$Hz, and those of TIC 461837097 are 5298 K and 249.0 $\mu$Hz. The two stars are the least evolved stars that show significant dipole-mode suppression. We demonstrate the two stars on the \teff$-$\Dnu{} diagram in Figure~\ref{fig:least} (here \Dnu{} is defined below eq.~\ref{eq:dnuT2}) with the \kepler{} subgiant discovered by \citet{Garcia2014AAp} as comparison. 
Findings on these subgiants suggest that the mode suppression can occur at late subgiant.

\section{Discussion and Conclusion} \label{sec:conclusion}

In this study, we conducted a comprehensive investigation of dipole mode suppression 8651 solar-like oscillators using data from TESS, and identified 179 stars exhibiting significant suppression (i.e. the spectra are fully resolved and have low dipole mode visibility). 
Our analysis reveals that dipole mode suppression becomes particularly prominent in stars with masses exceeding 1.4 $M_\odot$. This threshold aligns with previous findings from Kepler \citep{Stello2016Nature}, suggesting a shared underlying mechanism for dipole mode suppression in both subgiants and red giants.

Furthermore, we observed notable suppression in two subgiants (TIC 356735065 and TIC 260823248, see Fig.~\ref{fig:least}) and several early red giants near the base of the red-giant branch. These observations suggest that dipole mode suppression can begin before stars fully evolve onto the Red Giant Branch. Additionally, we found that the occurrence rate of dipole mode suppression is approximately 10\% in red giants (296 out of 2,532 stars) but only 2\% in subgiants (2 out of 101 stars).

While these preliminary results are clearly affected by small-number statistics, an increase in the suppression rate for more evolved stars is expected in the magnetic suppression hypothesis. As the stellar core contracts, its magnetic field will be amplified if the magnetic flux is conserved. Additionally, for the ``magnetic greenhouse effect'' to take place, the magnetic field must exceed a critical field strength, \( B_\mathrm{c} \). According to \citet{Fuller2015Science}, \( B_\mathrm{c} \) is proportional to the square of the oscillation frequency, which is roughly \numax{}. As stars evolve from the main sequence to the red giant branch, \numax{} decreases by more than an order of magnitude. Consequently, \( B_\mathrm{c} \) also decreases substantially. The combination of a strengthening central magnetic field and a decreasing critical field strength is expected to increase the rate of magnetic mode suppression as stars progress from the sub-giant phase to the giant phase. Future measurements will be needed to verify our tentative detection of a lower rate of dipole mode suppression in sub-giant stars.

This work provides a catalog of stars exhibiting significant or potential dipole mode suppression up to \numax{} = 670 $\mu\mathrm{Hz}$. Our sample serves as a valuable resource for future studies on this phenomenon, particularly those involving extended observation periods, higher cadence data, and stellar spectroscopy, which will be essential for deepening our understanding of dipole mode suppression and uncovering its underlying mechanisms.

\vspace{1cm}
\noindent
This work is supported by the National Natural Science Foundation of China (NSFC) grant 12373031, the Joint Research Fund in Astronomy (U2031203) under cooperative agreement between the National Natural Science Foundation of China (NSFC) and Chinese Academy of Sciences (CAS), and the NSFC grants 12090040, 12090042, and the CSST project. This work is partly supported by the National Science Foundation of China (Grant No. 12133005 to SM). This paper has also received funding from the European Research Council (ERC) under the European Union's Horizon 2020 research and innovation programme (CartographY GA. 804752). We also thank the \tess{} team for making this research possible.

\software{
emcee {\citep{emcee}},
PBjam {\citep{PBjam}}.
}


\bibliography{reference}{}

\begin{thebibliography}{}
\expandafter\ifx\csname natexlab\endcsname\relax\def\natexlab#1{#1}\fi
\providecommand{\url}[1]{\href{#1}{#1}}
\providecommand{\dodoi}[1]{doi:~\href{http://doi.org/#1}{\nolinkurl{#1}}}
\providecommand{\doeprint}[1]{\href{http://ascl.net/#1}{\nolinkurl{http://ascl.net/#1}}}
\providecommand{\doarXiv}[1]{\href{https://arxiv.org/abs/#1}{\nolinkurl{https://arxiv.org/abs/#1}}}

\bibitem[{{Ballot} {et~al.}(2011){Ballot}, {Barban, C.}, \& {Van\textquotesingle t Veer-Menneret, C.}}]{Ballot2011AAp}
{Ballot}, {Barban, C.}, \& {Van\textquotesingle t Veer-Menneret, C.} 2011, Astronomy \& Astrophysics (A\&A), 531, A124, \dodoi{10.1051/0004-6361/201016230}

\bibitem[{{Chaplin} {et~al.}(2014){Chaplin}, {Elsworth}, {Davies}, {Campante}, {Handberg}, {Miglio}, \& {Basu}}]{Chaplin2014MNRAS}
{Chaplin}, W.~J., {Elsworth}, Y., {Davies}, G.~R., {et~al.} 2014, \mnras, 445, 946, \dodoi{10.1093/mnras/stu1811}

\bibitem[{Chaplin {et~al.}(2013)Chaplin, Basu, Huber, Serenelli, Casagrande, Aguirre, Ball, Creevey, Gizon, Handberg, Karoff, Lutz, Marques, Miglio, Stello, Suran, Pricopi, Metcalfe, Monteiro, Molenda-Żakowicz, Appourchaux, Christensen-Dalsgaard, Elsworth, García, Houdek, Kjeldsen, Bonanno, Campante, Corsaro, Gaulme, Hekker, Mathur, Mosser, Régulo, \& Salabert}]{Chaplin2014APJS}
Chaplin, W.~J., Basu, S., Huber, D., {et~al.} 2013, The Astrophysical Journal Supplement Series, 210, 1, \dodoi{10.1088/0067-0049/210/1/1}

\bibitem[{{Donati} \& {Landstreet}(2009)}]{JFD2009ARAA}
{Donati}, J.~F., \& {Landstreet}, J.~D. 2009, Annual Review of Astronomy and Astrophysics, 47, 333, \dodoi{10.1146/annurev-astro-082708-101833}

\bibitem[{Ferrario {et~al.}(2015)Ferrario, Melatos, \& Zrake}]{Ferrario2015}
Ferrario, L., Melatos, A., \& Zrake, J. 2015, Space Science Reviews, 191, 77–109, \dodoi{10.1007/s11214-015-0138-y}

\bibitem[{{Foreman-Mackey} {et~al.}(2013){Foreman-Mackey}, {Hogg}, {Lang}, \& {Goodman}}]{emcee}
{Foreman-Mackey}, D., {Hogg}, D.~W., {Lang}, D., \& {Goodman}, J. 2013, \pasp, 125, 306, \dodoi{10.1086/670067}

\bibitem[{Fuller {et~al.}(2015)Fuller, Cantiello, Stello, Garcia, \& Bildsten}]{Fuller2015Science}
Fuller, J., Cantiello, M., Stello, D., Garcia, R.~A., \& Bildsten, L. 2015, Science, 350, 423–426, \dodoi{10.1126/science.aac6933}

\bibitem[{{Fuller} {et~al.}(2019){Fuller}, {Piro}, \& {Jermyn}}]{Fuller2019MNRAS}
{Fuller}, J., {Piro}, A.~L., \& {Jermyn}, A.~S. 2019, Monthly Notices of the Royal Astronomical Society, 485, 3661, \dodoi{10.1093/mnras/stz514}

\bibitem[{{Garc\'ia, R. A.} {et~al.}(2014){Garc\'ia, R. A.}, {Pérez Hernández, F.}, {Benomar, O.}, {Silva Aguirre, V.}, {Ballot, J.}, {Davies, G. R.}, {Doğan, G.}, {Stello, D.}, {Christensen-Dalsgaard, J.}, {Houdek, G.}, {Lignières, F.}, {Mathur, S.}, {Takata, M.}, {Ceillier, T.}, {Chaplin, W. J.}, {Mathis, S.}, {Mosser, B.}, {Ouazzani, R. M.}, {Pinsonneault, M. H.}, {Reese, D. R.}, {Régulo, C.}, {Salabert, D.}, {Thompson, M. J.}, {van Saders, J. L.}, {Neiner, C.}, \& {De Ridder, J.}}]{Garcia2014AAp}
{Garc\'ia, R. A.}, {Pérez Hernández, F.}, {Benomar, O.}, {et~al.} 2014, A\&A, 563, A84, \dodoi{10.1051/0004-6361/201322823}

\bibitem[{{Harvey}(1985)}]{Harvey_func}
{Harvey}, J. 1985, 235

\bibitem[{{Jenkins} \& {et al.}(2020)}]{Jenkins2020}
{Jenkins}, J.~M., \& {et al.} 2020, {Kepler Science Document KSCI-19081-003}, Kepler Science Document KSCI-19081-003, Edited by Jon M. Jenkins.

\bibitem[{{Kallinger} {et~al.}(2014){Kallinger}, {De Ridder}, {Hekker}, {Mathur}, {Mosser}, {Gruberbauer}, {Garc{\'\i}a}, {Karoff}, \& {Ballot}}]{Kallinger2014AAp}
{Kallinger}, T., {De Ridder}, J., {Hekker}, S., {et~al.} 2014, \aap, 570, A41, \dodoi{10.1051/0004-6361/201424313}

\bibitem[{Li {et~al.}(2017)Li, Bedding, Huber, Ball, Stello, Murphy, \& Bland-Hawthorn}]{Li2017MNRAS}
Li, T., Bedding, T.~R., Huber, D., {et~al.} 2017, Monthly Notices of the Royal Astronomical Society, 475, 981, \dodoi{10.1093/mnras/stx3079}

\bibitem[{Li {et~al.}(2023)Li, Bedding, Stello, Huber, Hon, Joyce, Li, Perkins, White, Zinn, Howard, Isaacson, Hey, \& Kjeldsen}]{Li2023MNRAS}
Li, Y., Bedding, T.~R., Stello, D., {et~al.} 2023, Monthly Notices of the Royal Astronomical Society, 523, 916, \dodoi{10.1093/mnras/stad1445}

\bibitem[{{Maeder} \& {Meynet}(2005)}]{Mag_dynamo}
{Maeder}, A., \& {Meynet}, G. 2005, \aap, 440, 1041, \dodoi{10.1051/0004-6361:20053261}

\bibitem[{{Mathur} {et~al.}(2011){Mathur}, {Hekker}, {Trampedach}, {Ballot}, {Kallinger}, {Buzasi}, {Garc{\'\i}a}, {Huber}, {Jim{\'e}nez}, {Mosser}, {Bedding}, {Elsworth}, {R{\'e}gulo}, {Stello}, {Chaplin}, {De Ridder}, {Hale}, {Kinemuchi}, {Kjeldsen}, {Mullally}, \& {Thompson}}]{Mathur2011APJ}
{Mathur}, S., {Hekker}, S., {Trampedach}, R., {et~al.} 2011, \apj, 741, 119, \dodoi{10.1088/0004-637X/741/2/119}

\bibitem[{Michel {et~al.}(2008)Michel, Baglin, Auvergne, Catala, Samadi, Baudin, Appourchaux, Barban, Weiss, Berthomieu, Boumier, Dupret, Garcia, Fridlund, Garrido, Goupil, Kjeldsen, Lebreton, Mosser, Grotsch-Noels, Janot-Pacheco, Provost, Roxburgh, Thoul, Toutain, Tiphène, Turck-Chieze, Vauclair, Vauclair, Aerts, Alecian, Ballot, Charpinet, Hubert, Lignières, Mathias, Monteiro, Neiner, Poretti, de~Medeiros, Ribas, Rieutord, Cortés, \& Zwintz}]{Michel2008Science}
Michel, E., Baglin, A., Auvergne, M., {et~al.} 2008, Science, 322, 558, \dodoi{10.1126/science.1163004}

\bibitem[{{Moss}(1982)}]{alpha-omega}
{Moss}, D. 1982, Monthly Notices of the Royal Astronomical Society, 201, 385, \dodoi{10.1093/mnras/201.2.385}

\bibitem[{{Mosser} {et~al.}(2015){Mosser}, {Vrard}, {Belkacem}, {Deheuvels}, \& {Goupil}}]{Mosser2015AA}
{Mosser}, B., {Vrard}, M., {Belkacem}, K., {Deheuvels}, S., \& {Goupil}, M.~J. 2015, \aap, 584, A50, \dodoi{10.1051/0004-6361/201527075}

\bibitem[{Mosser {et~al.}(2017)Mosser, Belkacem, Pinçon, Takata, Vrard, Barban, Goupil, Kallinger, \& Samadi}]{Mosser2017AA}
Mosser, B., Belkacem, K., Pinçon, C., {et~al.} 2017, Astronomy and Astrophysics, 598, A62, \dodoi{10.1051/0004-6361/201629494}

\bibitem[{{Mosser} {et~al.}(2017){Mosser}, {Belkacem}, {Pin{\c{c}}on}, {Takata}, {Vrard}, {Barban}, {Goupil}, {Kallinger}, \& {Samadi}}]{Mosser2016AAp}
{Mosser}, B., {Belkacem}, K., {Pin{\c{c}}on}, C., {et~al.} 2017, \aap, 598, A62, \dodoi{10.1051/0004-6361/201629494}

\bibitem[{{Murphy} {et~al.}(2013){Murphy}, {Shibahashi}, \& {Kurtz}}]{Murphy2013MNRAS}
{Murphy}, S.~J., {Shibahashi}, H., \& {Kurtz}, D.~W. 2013, \mnras, 430, 2986, \dodoi{10.1093/mnras/stt105}

\bibitem[{Nielsen {et~al.}(2021)Nielsen, Davies, Ball, Lyttle, Li~李坦, Hall, Chaplin, Gaulme, Carboneau, Ong~王加, García, Mosser, Roxburgh, Corsaro, Benomar, Moya, \& Lund}]{PBjam}
Nielsen, M.~B., Davies, G.~R., Ball, W.~H., {et~al.} 2021, The Astronomical Journal, 161, 62, \dodoi{10.3847/1538-3881/abcd39}

\bibitem[{Schneider {et~al.}(2019)Schneider, Ohlmann, Podsiadlowski, Röpke, Balbus, Pakmor, \& Springel}]{Schneider2019Nature}
Schneider, F. R.~N., Ohlmann, S.~T., Podsiadlowski, P., {et~al.} 2019, Nature, 574, 211–214, \dodoi{10.1038/s41586-019-1621-5}

\bibitem[{{Stello} {et~al.}(2016b){Stello}, {Cantiello}, {Fuller}, {Garcia}, \& {Huber}}]{Stello2016pasa}
{Stello}, D., {Cantiello}, M., {Fuller}, J., {Garcia}, R.~A., \& {Huber}, D. 2016b, \pasa, 33, e011, \dodoi{10.1017/pasa.2016.9}

\bibitem[{Stello {et~al.}(2016a)Stello, Cantiello, Fuller, Huber, García, Bedding, Bildsten, \& Aguirre}]{Stello2016Nature}
Stello, D., Cantiello, M., Fuller, J., {et~al.} 2016a, Nature, 529, 364–367, \dodoi{10.1038/nature16171}

\bibitem[{Twicken {et~al.}(2016)Twicken, Jenkins, Seader, Tenenbaum, Smith, Brownston, Burke, Catanzarite, Clarke, Cote, Girouard, Klaus, Li, McCauliff, Morris, Wohler, Campbell, Uddin, Zamudio, Sabale, Bryson, Caldwell, Christiansen, Coughlin, Haas, Henze, Sanderfer, \& Thompson}]{Twicken2016ANJ}
Twicken, J.~D., Jenkins, J.~M., Seader, S.~E., {et~al.} 2016, The Astronomical Journal, 152, 158, \dodoi{10.3847/0004-6256/152/6/158}

\bibitem[{{VanderPlas}(2018)}]{VanderPlas2018APJS}
{VanderPlas}, J.~T. 2018, \apjs, 236, 16, \dodoi{10.3847/1538-4365/aab766}

\bibitem[{{Yu} {et~al.}(2016){Yu}, {Huber}, {Bedding}, {Stello}, {Murphy}, {Xiang}, {Bi}, \& {Li}}]{Yu2016MNRAS}
{Yu}, J., {Huber}, D., {Bedding}, T.~R., {et~al.} 2016, \mnras, 463, 1297, \dodoi{10.1093/mnras/stw2074}

\bibitem[{Zhou {et~al.}(2024)Zhou, Bi, Yu, Li, Zhang, Li, Long, Li, Sun, \& Ye}]{Zhou2024APJ}
Zhou, J., Bi, S., Yu, J., {et~al.} 2024, The Astrophysical Journal Supplement Series, 271, 17, \dodoi{10.3847/1538-4365/ad18db}

\end{thebibliography}
\bibliographystyle{aasjournal}

\end{CJK*}
\end{document}